\documentstyle{mn}
\newcommand{\ale}{\ \raisebox{-.3ex}{$\stackrel{<}{\scriptstyle \sim}$}\ }
\newcommand{\age}{\ \raisebox{-.3ex}{$\stackrel{>}{\scriptstyle \sim}$}\ }

\newcommand{\vsc} {$d\,\Omega_{\rm orb}$}

\newcommand{\tsc} {$\Omega_{\rm orb}^{-1}$}

\input psfig

\title[Accretion discs in
intermediate polars]{The effects of tidally induced disc structure on
white dwarf accretion in intermediate polars}
	
\author[J.R. Murray et al.]{J.R. Murray$^1$,
	P.J. Armitage$^2$, L. Ferrario$^1$ and D.T. Wickramasinghe$^1$\\
	$^1$  
        The Astrophysical Theory Centre,
	Australian National University, ACT 0200, Australia \\
 	$^2$ Canadian Institute for Theoretical Astrophysics, McLennan Labs,
	60 St George St, Toronto, M5S 3H8, Canada}	
 	
\begin{document}
\maketitle
\begin{abstract}
We investigate the effects of tidally induced asymmetric disc structure on
accretion onto the white dwarf in intermediate polars. Using numerical
simulation, we show that it is possible for tidally induced spiral waves 
to propagate sufficiently far into the disc of an
intermediate polar that accretion onto the central white dwarf could
be modulated as a result. We suggest that accretion from the resulting
asymmetric inner disc may contribute to the observed X-ray and optical 
periodicities in the light curves of these systems. In contrast to 
the stream-fed accretion model for these periodicities, the tidal 
picture predicts that modulation can exist even for systems with 
weaker magnetic fields where the magnetospheric radius is smaller
than the radius of periastron of the mass transfer stream. 
We also predict that additional 
periodic components should exist in the emission from low mass ratio
intermediate polars displaying superhumps.
\end{abstract}


\section{Introduction}
In this paper we investigate the effects of tidally induced non-axisymmetric 
disc structure on white dwarf accretion in intermediate polars.

The largest tidal response in a close binary accretion disc is to the $m=2$
component of the binary potential. The response takes the form of a
pair of trailing spiral density waves. Excited at the outer edge of
the disc where tidal forces are strongest, these waves generally steepen to
become shocks as they extend inwards through the disc, and lead to 
a net outwards flux of angular momentum (Goldreich \& Tremaine 1979; 
Sawada, Matsuda \& Hachisu 1986). However in the cool discs of 
cataclysmic variables, where the Mach number of the Keplerian flow is 
probably ${\cal M}=v_\phi/c \age 25$, a variety of analytic and 
numerical arguments suggest that the efficiency of angular momentum 
transport is low, with a Shakura-Sunyaev (1973) $\alpha \sim 0.01$ or 
lower (Spruit 1987; Papaloizou \& Lin 1994; Papaloizou \& Lin 1995 and
references therein). In contrast, modelling of
dwarf novae outbursts suggests that high mass transfer cataclysmic variables 
have an $\alpha$ twenty to thirty times higher than this (Cannizzo 1993). 
The observation of a two armed spiral in the disc of the outbursting   
dwarf nova IP~Peg (Steeghs, Harlaftis \& Horne 1997) does not appear 
to alter this general conclusion (Armitage \& Murray 1998). 

In this paper, we suggest that spiral waves in close binary
could have interesting observational consequences even if 
they are {\em unable} to account for the bulk of the disc angular 
momentum transport. The presence of spiral waves breaks the
axisymmetry of the inner disc and tells the accreting
star the orbital phase of its companion, resulting in an additional
orbital period dependent
variation in the accretion rate onto the white dwarf. The amplitude 
of such variations will depend on the strength of the spiral waves.
In the presence of turbulent fluid motions or vertical temperature gradients 
in the disc, we expect that waves will be damped (Lubow \& Ogilvie 1998), 
perhaps strongly. Intermediate polars, where the inner disc radius is 
large and the disc's thermal structure is modified by irradiation, may 
therefore be the best place to search for these effects. 

For systems with mass ratios $q \age 0.25$, which encompasses most 
intermediate polars, the tidally induced spiral structure is fixed
in the binary frame. In binaries with a more extreme mass ratio, 
$q \la 0.25$, an additional complication is the possibility of
resonance between the binary orbit and gas orbits in the outer
disc. This can force the disc into an eccentric precessing state 
(Whitehurst 1988, Hirose \& Osaki 1990, Lubow 1991, Murray 1998), 
observationally identified with the superhumps seen in SU~UMa stars
in superoutburst, and in some short period novalikes. We also
investigate the effects of this resonant tidal disc structure
on accretion in intermediate polars.

In section 2 we briefly summarise the periodicities that are found in
the optical and X-ray light curves of intermediate polars, and comment
on the strengths and weaknesses of the stream overflow model. In
section 3 we describe the details of our numerical method. Simulations
of tidally induced spiral waves are shown in section 4. A simulation
of superhumps in an intermediate polar is outlined in section 5, and
our conclusions are presented in section 6.

\section{Periodicities in the light curves of Intermediate Polars}
In intermediate polar systems, the magnetic field of the white dwarf
is sufficiently strong as to truncate or even completely disrupt the
surrounding accretion disc, but not so strong as to synchronise the
spin of the white dwarf with the rotation of the binary. In those
systems where it is not completely disrupted, the accretion disc is
typically thought to extend down to radii 
$r_{\rm in} \simeq 8-12\: r_{\rm wd}$. Inwards of this radius, 
the motion of the gas is dominated by the magnetic field.

For comprehensive reviews of the state of knowledge regarding
intermediate polars, the reader is referred to Patterson (1994) and 
Warner (1995). Of particular interest here is the range of
periodicities that are apparent in both the optical and X-ray emission
from these systems. For example, Buckley \& Tuohy (1989) measured the
orbital period and white dwarf spin period of TX Columbae to be
$5.72 \pm 0.07$ hours and $1911 \pm 2$ seconds respectively. They
found the optical emission to be modulated at frequencies $\omega_{\rm
spin}-\Omega_{\rm orb}$, $\omega_{\rm spin}-2\,\Omega_{\rm orb}$, and
$\omega_{\rm spin}+\,\Omega_{\rm orb}$, and the X-ray
emission to be modulated at frequencies  $\omega_{\rm spin}$, 
and  $\omega_{\rm spin}-\Omega_{\rm orb}$.
Norton et al. (1997) found the relative strengths of the X-ray
periodicities changed over a time scale of one year. In 1994 they
observed TX~Col with ASCA and found the strongest modulation of the
X-ray emission occurred at $\omega_{\rm spin}$ and $\Omega_{\rm
orb}$. Using ROSAT one year later, they found the system also
displayed modulation at $\omega_{\rm spin}-\Omega_{\rm orb}$ and 
$\omega_{\rm spin}-2\,\Omega_{\rm orb}$. The X-ray light curve of FO
Aquarii is similarly variable (Beardmore et al. 1998).

The presence of these different frequencies in the light curve
provides clues as to the mode of accretion. Modulation at a frequency 
of $\omega_{\rm spin}$ implies that accretion onto the white
dwarf occurs at a roughly fixed location on the surface, most 
likely due to misaligned spin and magnetic axes, while the presence
of a frequency $\Omega_{\rm orb}$ implies structure fixed in 
the rotating binary frame. The mixed frequencies such as
$\omega_{\rm spin}-\Omega_{\rm orb}$ then suggest that there
is asymmetry in the inner disc, where the flow is becoming 
magnetically dominated by the white dwarf field, that is aware
of the phase of the binary companion. 

The stream-fed accretion model provides one way of creating these 
conditions. In this model there are two components to the accretion 
onto the white dwarf; flow from the inner edge of an axisymmetric
disc, and direct accretion from an overflowing stream that carries 
some fraction of the mass transfer from the inner Lagrange point.
The stream trajectory is essentially ballistic and fixed in the 
binary frame, so the accretion rate from this component, $\dot m_{\rm s}$,  
will be modulated at the beat between the spin period of the white dwarf 
and the orbital period of the binary. Norton et al. (1997) proposed
such a model to explain their observations of TX Col, and similar 
ideas have been suggested both for EX Hya in outburst 
(Hellier et al. 1989), and to explain the empirically based 
$10:1$ ratio between orbital and spin periods (Warner \&
Wickramasinghe 1991, King \& Lasota 1991).
There are convincing observational (Marsh \& Horne 1989), analytical 
(Lubow 1989) and numerical (Armitage \& Livio 1998; Blondin 1998)
arguments supporting the basic idea that a significant fraction of 
the gas stream from the secondary skims over the outer edge of the 
disc only to impact upon the disc plane at smaller radii. 

Despite this, the circumstances in which stream overflow 
can give rise to modulated accretion are limited. 
In this model, the radius of periastron, $r_{\rm per}$, achieved by a
ballistic particle emerging from the inner Lagrange point, 
is the minimum radius to which an overflowing stream can
penetrate. Any encounter with the disc,
either at the hot spot or along the disc surface, will result in gas
from the stream being entrained in the disc flow at radii $r > r_{\rm
per}$.
 Therefore if the magnetospheric radius
$r_{\rm m} < r_{\rm per}$, accretion is expected to occur
from an axisymmetric inner disc. 
Conversely, if $r_{\rm m} \age r_{\rm per}$, then it is unclear
whether a disc could have formed in the first place, though once 
formed (perhaps when the accretion rate was high and $r_{\rm m}$ 
was smaller) it might well be able to survive. The long term
behaviour of systems in this regime is, frankly, extremely 
murky, especially when the possibility of significant torques 
applied at the inner boundary is included (King \& Lasota 1991). 
However, assuming that there {\em is} a combination of disc and 
stream fed accretion, the clear prediction of the stream fed model 
is that the modulation should be strongest when the accretion rate
is low. In this state the magnetospheric radius is largest, and the 
fraction of material that overflows in the stream is probably enhanced
(Armitage \& Livio 1998), both of which favour a larger pulsed fraction 
of the emission at the orbital side band period $\omega-\Omega_{\rm
orb}$. At least in EX Hya, the observed periodicity was
most apparent when the system was in outburst (Hellier et al. 1989) 
and $r_{\rm m} $ would
likely be at a minimum.

In summary then, it appears necessary for the adequate explanation of
the observed modulations in intermediate polar light curves, 
that accretion onto the white dwarf be modulated by the rotational motion of
the white dwarf {\em with respect to the binary frame}. 
Simple orbital variation of the net system luminosity due to
asymmetries in the reprocessed radiation from  sites such as the
secondary, and the disc bright spot, cannot explain all the
frequencies seen in these systems. Although direct accretion from the
$L_{\rm 1}$ stream would naturally give a modulated rate of accretion,
there are arguments to suggest that this is not the whole picture.

\section{Numerical Method}
The two and three dimensional simulations described in this paper 
were completed using a Smooth Particle Hydrodynamics (SPH) code that 
has been described in detail elsewhere (Murray 1996a). 
Other calculations made using the code can be found in
Murray (1998), Armitage \& Murray (1998), and Murray \& Armitage
(1998). The SPH technique is comprehensively reviewed in Monaghan (1992).

Murray (1996b), and Yukawa, Boffin \& Sawada
(1997) presented SPH calculations of spiral shocks in close binary
accretion discs. Previously it had been thought that the artificial
shock technique used for improving shock capture in SPH simulations
was too dispersive for meaningful results to be obtained in shearing
discs. In fact,  previous SPH simulations that had failed to show
nonlinear spiral waves (e.g. Murray 1996a), had done so because 
high values for the Mach number $\cal{M}$ were used, and the $m=2$ tidal
response of the disc was tightly wrapped and rapidly damped. Simulations 
shown in the next section will reinforce this point.

SPH codes incorporate a variety of different viscous terms in the 
equations of motion. These terms are generally used both to ensure stable 
resolution of shocks, and to model the anomalously high shear viscosity 
that occurs in discs. The simulations described here follow
Meglicki, Wickramasinghe \& Bicknell (1993) and 
Murray (1996a), and use a dissipation  term based upon the {\em
linear} artificial 
viscosity term described in Monaghan (1992). However in the shearing flow of 
an accretion disc it is appropriate that viscous forces be
enabled for receding particles as well as for approaching
particles, so the {\em viscous switch} is disabled. 
In the continuum limit, the introduction of such a term is
equivalent to introducing shear and bulk viscosity, in a fixed ratio,
into the fluid equation of motion (see equation 3, Murray 1998).

Flebbe et al. (1994) and Watkins et al. (1996) described terms that
introduce independent shear and bulk viscosities into the equations of
motion. Kunze, Speith \& Riffert (1997) describe several two
dimensional simulations made with the Flebbe et al. formulation. As
has been previously pointed out, the forces these terms
introduce between particles are not antisymmetric and along the
particles' line of centres, so angular momentum is not conserved at
the particle level. In fact, Riffert et al. (1995) show that, with the
Flebbe et al. term, total angular angular momentum is only {\em
approximately} conserved (i.e. to second order in the smoothing length).

In their disc simulations Yukawa et al. incorporated both the linear
($\alpha$) and nonlinear ($\beta$)  SPH artificial viscosity terms,
exactly as described in Monaghan (1992). Tests show that the 
nonlinear term is often necessary to prevent particle interpenetration
in high Mach number shocks, particularly with an isothermal equation 
of state (e.g. Bate 1995). We have found that in disc situations, the linear 
term is sufficient for adequate shock capture. Furthermore, in simulations 
we completed with a non-zero $\beta$, many particles from the outer
regions of the disc were flung to very large radii. As a result the
calculations  slowed down enormously, and large numbers of particles
either escaped the system or were returned to the secondary. We  found
that the use of the nonlinear viscous term lead to a degraded treatment of
disc boundaries, and to less reliable estimates of accretion rates on
to the central object.

In these simulations, masses are scaled to the total system mass,
$M=M_1+M_2$, lengths are scaled to the  interstellar separation, $d$,
and times are scaled to the reciprocal of the orbital angular velocity,
\tsc. We choose the centre of mass of the binary to be the origin of
our inertial coordinate system.

\begin{figure*}
%\mbox{\psfig{figure=d6lr.eps,width=15.5cm}}
\mbox{\psfig{figure=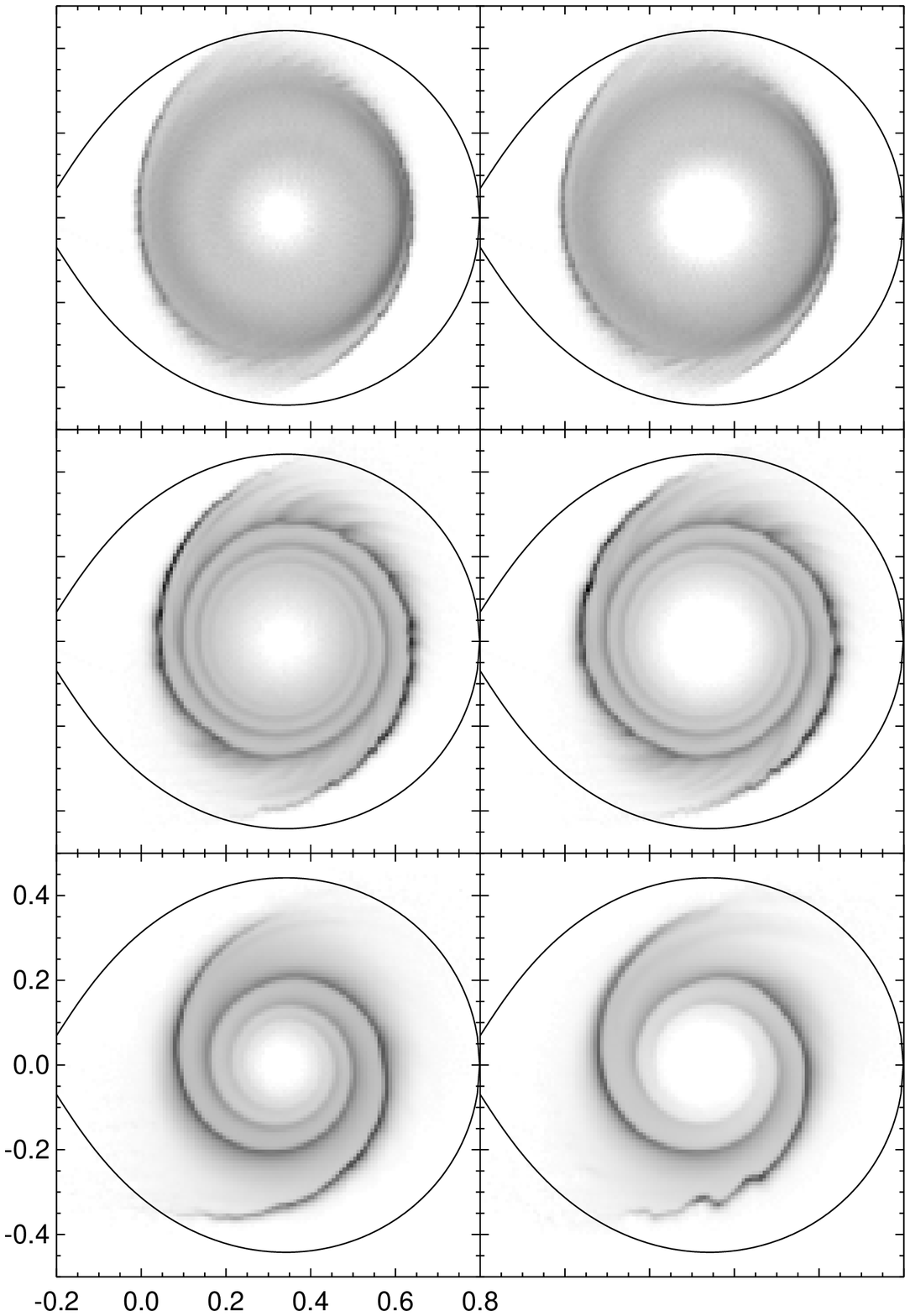,width=15.cm}}
\caption{The dependence of spiral structure upon disc temperature
and upon the radius of the inner boundary $r_{\rm in}$.
Grey scale density maps for six disc calculations at time 
$t=20.00$ \tsc.
The same density scale is used in all frames.
The SPH artificial viscosity parameter
$\zeta=1.0$ in all simulations. $M_1=1.02 M_\odot$, $M_2=0.5
M_\odot$.  
$r_{\rm in}= 0.05\,\, d$ in the
simulations on  the  left hand side, and  $r_{\rm in}= 0.10 \,\, d$ in
the right hand  calculations.  The 
sound speed is $c_0  (2\,r)^{-3/8}$, where
$c_0=0.02$ \vsc\, in the top two simulations, $c_0=0.05$ \vsc\, 
in the middle two panels and 
 $c_0=0.10$ \vsc\, for the bottom two calculations. The Roche lobe of the
primary is shown as a solid line in each panel. Coordinatess (marked on
both axes) are centred on the binary centre of mass and are 
scaled to the interstellar separation $d$.}
\label{fig:sixdens}
\end{figure*}

\begin{figure*}
%\mbox{\psfig{figure=bltn.eps,width=15.5cm}}
\mbox{\psfig{figure=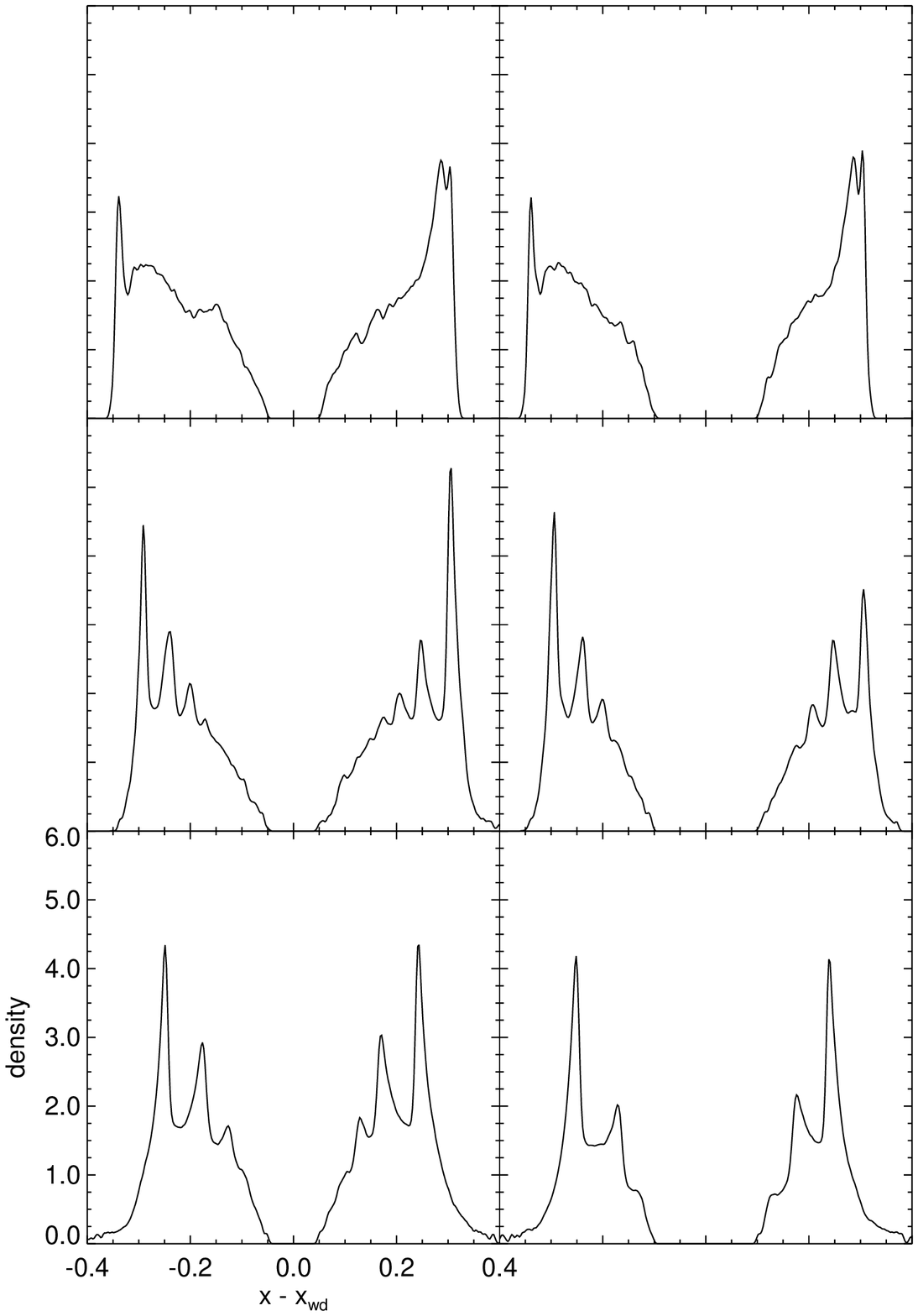,width=15.cm}}
\caption{For each of the six discs shown in Fig.~\ref{fig:sixdens},
the surface density along the line $y=0$ (the binary axis) at time
$t=20.00$ \tsc\, is plotted. 
The same length (horizontal axis) and density (vertical axis)
scalings are used in each panel. The position is measured with respect
to the centre of mass of the white dwarf ($x_{\rm wd}$).}
\label{fig:dcut6}
\end{figure*}

\section{Spiral Wave Simulations}
In this section we describe simulations  that show that it is possible for 
tidally induced spiral waves to propagate sufficiently far into the 
disc of an intermediate polar that the white dwarf accretion may be 
modulated on the beat period between the white dwarf spin and binary 
orbital periods.

As we are interested in the propagation of spiral structure to small radii, 
we elected  to complete several high resolution two dimensional simulations 
as opposed to lower resolution calculations in three dimensions. We use a 
constant smoothing length $h=0.005\, d$, where $d$ is the stellar 
separation. 

A Keplerian disc with a surface density profile $\Sigma = {\rm constant}$
was constructed by laying down particles on concentric rings
$\frac{1}{2}\,h$ apart, out to a radius $r=0.395\,d$. Particle spacing
within each ring was also $\frac{1}{2}\,h$. Each simulation began with
$74 445$ particles. Particles were then added  at the $L_1$
point in the  manner described  in Murray (1998), at a
rate of one particle per $0.05$ \tsc. We take a representative 
value for the binary mass ratio of $q=0.5$. 

Present theoretical uncertainty precludes any attempt to model the details 
of the interaction between the inner disc and the magnetic field of the white
dwarf. We simply use an open boundary condition, so that at every
time step particles with a radius $r<r_{\rm in}$ are removed from the
calculation (as are particles that either return to the secondary or
that are further from the primary than is the $L_1$
point). This boundary condition neglects the influence of magnetic 
torques on the disc exterior to $r_{\rm m}$, which may be important 
in some systems (Wynn \& King 1995).

An $r^{-3/4}$ temperature profile is imposed upon the disc by requiring
that the sound speed
\begin{equation}
c =\,c_0 \:(\frac{r}{r_0})^{-3/8},
\end{equation}
where $c_0$ is the sound speed at some arbitrary reference radius $r_0$, 
which we take as $r_0=0.5\,d$.

With the parameters given above, 
the Mach number of the Keplerian flow,
\begin{equation}
{\cal M}(r)\simeq 1.06\,\,{c_0}^{-1}\,r^{-1/8}.
\end{equation}
We complete simulations for $c_0=0.02,0.05,0.10$ \vsc, giving
Mach numbers of 60, 24 and 12 respectively  at the approximate outer
edge of the disc, $r=0.4\,d$. For a
Shakura-Sunyaev disc solution (Shakura \& Sunyaev 1973), ${\cal M}
\simeq 25$ is roughly appropriate for a CV disc.
We note that the Yukawa et al. calculations assumed a hotter disc,  
with ${\cal M} \simeq 5$.

We have previously shown (Murray 1996a) that the SPH linear artifical
viscosity term generates a shear viscosity equivalent to that of a disc
with Shakura-Sunyaev viscosity parameter
\begin{equation}
\alpha(r)=\frac{1}{8}\,\zeta\,\Omega(r)\,c^{-1}\,h.
\label{eq:zetatoalpha}
\end{equation}
Here $\zeta$ is the SPH artificial viscosity parameter. 
Although the discs described here are noticeably asymmetric, the
velocity divergence is only significant in the outer regions ($r \ga
0.3\,d$). In these outer regions the SPH artificial viscosity also
generates a bulk viscosity which cannot be represented in terms of an
$\alpha$ parameter.
With $\zeta=1.0$, and the parameter values listed above,
\begin{equation}
\alpha(r)\simeq\,0.13\,\frac{h}{c_0}\,r^{-9/8}.
\end{equation}
For sound speed $c_0=0.05$ \vsc, $\alpha(0.4)\simeq 0.04$ and
$\alpha(0.1)\simeq 0.2$. 

As we have already mentioned, mass transfer from the secondary stars
was included in the simulations. However the computational
time required for these calculations is so large that it was not
feasible to evolve each disc to steady state. Instead we ran each
calculation to a point where we were certain that the discs' {\em tidal
structure was stationary} i.e. until the location and pitch angle of
the spiral arms had stabilised. This was largely the case by $t=20$
\tsc, although most simulations were actually run for  $60$ or $100$ \tsc.

The surface density  of six two dimensional simulations at
time $t=20$ \tsc\, are shown
in Fig.~\ref{fig:sixdens}. Each panel shows the disc and the
primary's Roche lobe in a frame that co-rotates with the binary such
that the origin is at the binary's centre of mass, and the primary
lies along the positive x-axis.
The grey scale maps economically demonstrate a point that is already 
well known; that in cool viscous discs (top two panels) the spiral 
response is weak and restricted to the outermost regions. For discs 
of intermediate Mach number (and value of $\alpha$), tightly wrapped
spiral waves extend down to radii $r < 0.2d$ before being completely 
damped. For yet higher temperatures ($c_0 = 0.10 \,d \Omega_{\rm orb}$), 
the spiral waves can easily penetrate all the way to the inner edge 
of the disc.

Simulations were completed with $r_{\rm in}=0.05\,d$ (left hand
panels), and with $r_{\rm in}=0.10\,d$. Fig.~\ref{fig:sixdens} 
clearly shows that 
the position of the inner boundary has minimal effect upon the disc's 
tidal structure. 

Fig.~\ref{fig:dcut6} shows the surface density
along the line $y=0$ for each of the density maps shown in
Fig.~\ref{fig:sixdens}.  We caution that although the 
tidal structure of each disc {\em was} stationary,
the disc masses have not reached stable values.
Rather more accretion and radial spreading has occurred in the
hottest discs, and so the density  in these simulations is reduced
with respect to the cooler calculations. However it is clear 
that except for the coolest disc, the spiral waves maintain 
significant overdensities down to close to the inner 
boundary. 

We also wish to compare the two spiral waves within a given
simulation. Should the two waves differ significantly in strength in
the inner disc, then we would expect matter to be accreted
preferentially from the stronger of the two waves. The relative
strengths of the resulting periodic luminosity variations would be
different than if the arms had equal strength. Although
Fig.~\ref{fig:dcut6} reveals the two arms to be significantly
different in the outer disc, in the inner disc the arms are very
similar in strength and in the location at which they intersect the
x-axis. In other words, these simulations do not reveal any
significant differences between the two spiral waves {\em in the inner
regions of the disc}. However more detailed simulations may be
required to definitively answer this question.

The open inner boundary strongly influences the simulations in an
annulus $r_{\rm in} < r \ale r_{\rm in} + 0.05\,d$. The effective
inner boundary of the calculations should be taken at 
$r\simeq  r_{\rm in} + 0.05\,d$.  Fig.~\ref{fig:dcut6} shows again,
that whilst tidal structure is restricted to the outer regions in the
coolest discs, both the intermediate temperature and hot disc
simulations show spiral structure extending down to radii $r \ale 0.20
d$. The key result is that, in contrast to stream overflow, there is
no minimum radius below which spiral structure is forbidden. The
degree to which spiral waves affect flow in the inner regions depends
upon the temperature, and upon the rate at which the waves are damped. We
will return to this point in the discussion section.
 
\section{Superhumps}
The majority of intermediate polars lie above the period gap, implying
their
mass ratios are too large for the eccentric resonance to be populated by disc
material. There are some exceptions however. Most notably EX Hya has
an orbital period $P_{\rm orb}=0.068234$ days, 
and an estimated mass
ratio $q\simeq 1/6$ (Hellier et al. 1987). 
One would expect disc precession to occur  in this system. 
Superhumps have in fact been
observed in the intermediate polar candidate, VZ Pyxidis (Kato \&
Nogami, 1997). Most recently, the X-ray
source RX J0757.0+6306 has been identified as an intermediate polar
with orbital
and white dwarf spin periods of $81 \pm 5$ and $8.52 \pm 0.15$ minutes
respectively (Tovmaissian et al. 1998).

To investigate the properties of superhumps in intermediate polars
we completed a three dimensional SPH simulation of a disc in a binary 
with mass ratio $q=3/17$. To set up an eccentric precessing disc,
we began the calculation from zero disc mass, and
added single particles at regular time intervals ($\Delta t=0.01$
\tsc), in circular orbit at the circularisation radius. The smoothing
length was kept fixed at $h=0.02 \,d$, and the sound speed $c=0.02$
\vsc. In fact with these values, the smoothing
length was everywhere greater than the pressure scale height $H$, 
so that the vertical structure was not resolved and the
calculation was in effect two dimensional. Neighbour numbers
averaged $160$ for the calculation.

As in the previous section, an open inner boundary condition was 
used. Initially we set the radius of the inner boundary $r_{\rm in}=0.02\,d$. 
We then followed the simulation until the disc had encountered the resonance 
and subsequently reached an eccentric equilibrium state.
The disc precession period
agreed with the results of previous two dimensional simulations
(Murray 1996a, 1998).

We set our SPH artificial viscosity parameter $\zeta=1.0$. For the 
{\em three dimensional} SPH code, 
\begin{equation}
\nu=\frac{1}{10}\,\zeta c\, h, 
\label{eq:tdv}
\end{equation}
so for this isothermal run the disc had a shear viscosity $\nu$ that 
is constant with radius, i.e. the Shakura-Sunyaev parameter 
$\alpha \propto r^{-3/2}$. 

At time $t=644.00$ \tsc\, the radius of the inner boundary was adjusted 
to a value more appropriate to an intermediate polar. Particles lying interior 
to the new inner boundary,  $r_{\rm in}=0.10\,d$, were immediately rejected 
from the simulation. The calculation restarted with $18094$ particles in the 
disc. At this stage we changed the mode of mass addition. Particles were now 
added (at the same rate as before) at the $L_1$ point, rather than at
$r_{\rm circ}$. In the initial calculation mass was added at $r_{\rm
circ}$ so that the disc could rapidly reach an equilibrium with the
resonance. But adding mass in the inner disc also artificially
symmetrises the flow there. Any dependance on the disc precession
would thus be washed out of the accretion from the disc's inner edge.
The run was then continued for a further $366.00$ \tsc.

\begin{figure}
\mbox{\psfig{figure=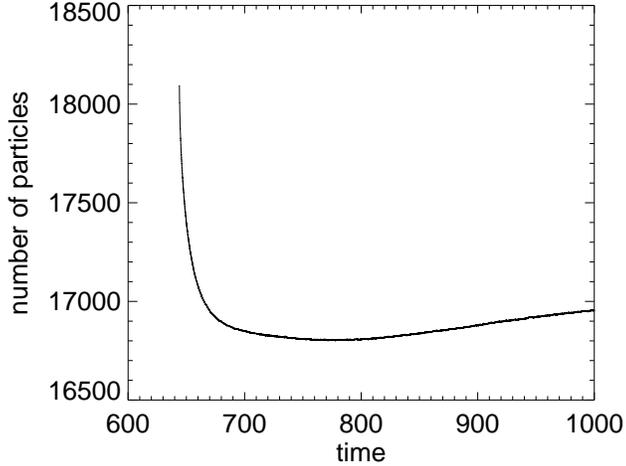,height=7.cm}}
\caption{Disc particle number (which is proportional to the  disc mass)
as a function of time for the superhump simulation. Particle number was
recorded every $\Delta t=0.01$
\tsc.}
\label{fig:np}
\end{figure}

Fig.~\ref{fig:np} shows the disc mass as a function of time for the 
``intermediate polar'' section of the simulation. 
The disc took approximately 100 \tsc\, 
to accomodate the new inner boundary condition and the new mode of mass 
addition. A minimum of 16799 particles was reached at time $t=771.95$ \tsc. 
Thereafter the disc slowly accumulated mass, so that by the end 
of the calculation it contained $16955$ particles.

No particles were removed from the simulation as a result of either
escaping to large radii or returning to the secondary, and mass addition
occurred at a constant rate. Therefore the rate of accretion onto the
white dwarf can be directly determined from the change in the disc mass.

\begin{figure}
\mbox{\psfig{figure=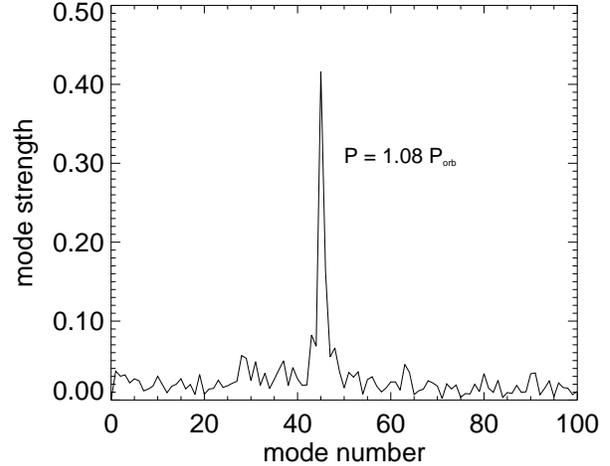,height=7.cm}}
\caption{Fourier power spectrum of the disc mass time
series. Both the trend and short time scale variability were removed
from the time series before the Fourier transform was taken. The
time series covered a period $305.56$ \tsc.}
\label{fig:pws}
\end{figure}

\begin{figure}
\mbox{\psfig{figure=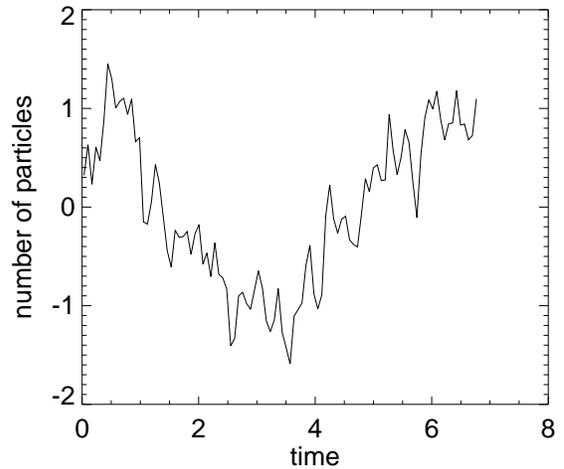,height=7.cm}}
\caption{Number of particles in the disc folded on $P_{\rm acc}$
(after removal of the mean and long term trend) and
averaged in 100 time bins.}
\label{fig:shfold}
\end{figure}

After removing the trend and  the shortest
time scale variability, we took the Fourier transform of the disc mass
time series to reveal an underlying periodicity in the white dwarf accretion
rate. The power spectrum
(Fig.~\ref{fig:pws}) has only one significant peak, at 45 cycles,
which corresponds to a period $P_{\rm acc}=1.08 \pm 0.02\,P_{\rm orb}$.
When folded on this period, the disc mass shows a noisy but highly 
significant periodicity (Fig.~\ref{fig:shfold}). The disc
precession period {\em obtained from the simulation light curve}
$P_{\rm sh}=1.07 \pm 0.02\,P_{\rm orb}$. 
Thus we found accretion from the inner edge of the {\em truncated} disc
was modulated on the period of the disc's motion {\em as measured in
the binary frame}. In the eccentric disc model for superhumps this is
the superhump period, $P_{\rm sh}$.

What are the observational consequences of this result? 
In Murray (1998) it was demonstrated that the modulated component of
a precessing disc's luminosity emerged from a disc region that was
{\em fixed in the binary frame}.
On this basis we similarly expect that the modulated component
of the accretion comes from a point in the disc that is fixed in
the binary frame. Any emission resultant from the accretion onto
the  white dwarf will be modulated on the beat between the white dwarf
spin {\em as measured in the binary frame}, and the superhump
period. Thus we expect to see modulations in the X-ray emission from
such systems at frequencies 
\begin{equation}
(\omega_{\rm spin} - \Omega_{\rm orb}) -
\omega_{\rm sh} \simeq \omega_{\rm spin} - 2\,\Omega_{\rm orb}, 
\end{equation}
and
\begin{equation}
(\omega_{\rm spin} - \Omega_{\rm orb}) + \omega_{\rm sh} \simeq
\omega_{\rm spin}(1 - \delta), 
\end{equation}
where $\delta \ll 1$. 

\section{Discussion}
In this paper we have presented simulations of spiral structure in the 
accretion discs of intermediate polars.  Our results suggest that the 
observed periodic emission in intermediate polars could be caused by the 
interaction of the white dwarf magnetic field with asymmetric structure 
in the accretion disc induced by the tidal field of the binary. In contrast 
to the stream overflow model for these periodicites, the tidal model predicts 
that modulation can occur even if $r_{\rm m} < r_{\rm per}$, and that 
frequencies related to the superhump period should be present 
in low mass ratio systems ($q \ale 0.25$) displaying superhumps. 

\begin{table}
\caption{Methods of introducing various frequencies in the light curves
of magnetic cataclysmic variables.}
\label{tbl:freq}
\begin{tabular}{p{5.cm}c}
Mechanism & Expected Frequency \\
\\
Accretion in Magnetic CVs with Discs\\
from two spiral arms to one pole & $\omega_{\rm
spin}-2\Omega_{\rm orb}$\\
from two spiral arms to two poles & $2\omega_{\rm spin}-2\Omega_{\rm orb}$\\
from eccentric disc onto one pole & $\omega_{\rm
spin}-\Omega_{\rm orb}\pm\Omega_{\rm sh}$\\
\\
Accretion in Discless Magnetic CVs\\
from stream to one pole & $\omega_{\rm spin}-\Omega_{\rm orb}$\\
from stream to two poles & $2\omega_{\rm spin}-\Omega_{\rm
orb}$\\
\\
Radiation from one pole reprocessed\\
at disc hot spot & 
$\omega_{\rm spin}-\Omega_{\rm orb}$\\
by spiral arms & 
$\omega_{\rm spin}-2\Omega_{\rm orb}$\\
\end{tabular}
\end{table}

In Table~\ref{tbl:freq} we have summarised the frequencies
that each mechanism would be expected to introduce into a magnetic
cataclysmic variable's light curve. In addition, accretion from a disc
is expected at some level to lead to 
modulation at the
spin period $\omega_{\rm spin}$.
Wynn \& King (1992) have
demonstrated that the X-ray power spectrum generated by stream
accretion is strongly dependant upon
the configuration of the white dwarf's magnetic field, and upon the
observer's inclination to the binary orbital plane. Therefore we caution against
using this table in isolation to determine the mode of
accretion occurring in a particular system. However, consider the
recently discovered  intermediate polar RX~J1238-38.  Optical
photometry by Buckley et al. (1998) revealed periodicities at 1860 and
2147 seconds, {\em ``with similar amplitudes of $\simeq 8 \%$''}. Spectroscopy
showed emission lines varied on the longer of the two periods,
indicating that the white dwarf spin period was $2147$ s. Further
analysis of the emission lines revealed evidence of a longer period at
$5300^{+1260}_{-850}$ s. Buckley et al. found the $1860$ second period
was consistent with a frequency $2(\omega_{\rm spin} - \Omega_{\rm
orb})$. Now in the stream-accretion picture such a frequency is not
expected to feature prominently in the power spectrum. On the other
hand, accretion from a tidally distorted disc  would naturally produce
this periodicity. 

Of necessity, we have had to gloss over the uncertain details of the 
coupling between the white dwarf magnetic field and the inner disc. This 
could be an important omission, especially for white dwarfs whose spin rate 
and magnetic field are such that the coupling exerts a positive torque 
on the inner disc immediately outside the magnetospheric radius. By 
analogue with the results of Artymowicz \& Lubow (1996) for 
{\em circumbinary} discs (where the torques are gravitational), such 
a situation could lead to an additional sources of asymmetry in the 
inner disc, and associated modulation of the accretion rate. 

We chose to simplify the thermodynamics, by prescribing
the radial temperature profile of each disc. Whilst this assumption
influences the detail of the simulations (e.g. larger density contrasts
can occur across our essentially isothermal shocks than if we had used
an adiabatic equation of state), it does not take away from the
general result that the spiral modes are stronger in hotter discs.
Our results  are consistent with the numerical and analytic work of
other authors (see e.g. the simulations using a polytropic equation
of state completed by R\'o\.zyczka and Spruit 1989), though the 
possible effects of vertical temperature gradients (Lubow \& Ogilvie 1998) 
have not been addressed systematically in any simulations to date.   

Fig.~\ref{fig:sixdens} shows that an increase in the disc
temperature results in a strengthening of the spiral modes.
This would
suggest an increase in the modulated component of the accretion onto
the white dwarf. On the other hand, an increase in disc temperature
implies an increase in the mass flux through the disc $\dot M_{\rm
d}$. Under these circumstances we would expect $r_{\rm in}$ to
decrease which in turn would make it more difficult for the spiral
waves to penetrate right to the disc's inner edge. Worse, the 
damping due to turbulence in the disc is also likely to vary with 
the accretion rate. Assessing the relative importance of these
effects is difficult. However it is natural to expect the modulated
fraction to change. Thus changes in disc temperature, as may be caused
by changes in the mass flux from the secondary, could explain the
finding of Norton et al.  (1997) that the relative strengths of the
periodicities in the light curve of TX Col changed over the time scale
of one year.

Theoretical work suggests that the development of observable spiral 
structure that extends over a large range in radii depends on the 
disc Mach number, the presence of vertical temperature gradients, 
and the strength of any turbulent viscosity (Savonije, Papaloizou \& 
Lin 1994; Godon 1997; Lubow \& Ogilvie 1998). Doppler tomography of a variety 
of systems can potentially help in understanding these mechanisms. 
For example, spiral waves that were present in the inner disc of 
quiescent dwarf novae, but not present in outburst, would suggest that 
damping due to turbulence was the dominant process, the converse would 
imply that the Mach number was of greater import. Further calculations 
are also required in order to quantify the relative importance of these 
effects.

Recently, Steeghs et al. (1997) used Doppler tomographic
techniques to identify spiral structure in the accretion disc of
IP~Peg down to radii $r \simeq 0.28\,d$. Similar observational
identification of spiral structure in intermediate polar systems would 
provide strong support for the scenario presented here.

\section*{Acknowledgments}
The spiral wave
computations were completed using the {\small SGI} Power Challenge
computer at The Australian National University Supercomputer Facility.


\begin{thebibliography}{}
\bibitem{philandmario}
Armitage, P.J., Livio, M., 1998, ApJ, 493, 898
\bibitem{philandme}
Armitage, P.J., Murray, J.R., 1998, MNRAS, 297, L81
\bibitem{artymowicz}
Artymowicz, P., Lubow, S.H., 1996, ApJ, 467, L77
\bibitem{bate}
Bate, M., 1995, Ph.D. Thesis, University of Cambridge, p. 43
\bibitem{}
Beardmore, A.P., Mukai, K., Norton, A.J., Osborne, J.P., Hellier, C.,
1998, MNRAS, 297, 337
\bibitem{} 
Blondin, J.M., 1998, in ``Accretion Processes in Astrophysical Systems: 
Some Like it Hot!'', eds S.S. Holt \& T.R. Kallman, AIP Conf. Proc. 431, 
p.~309
\bibitem{buckley}
Buckley, D.A.H., Tuohy, I.R., 1989, ApJ, 344, 376
\bibitem{BCRW}
Buckley, D.A.H., Cropper, M., Ramsay, G., Wickramasinghe, D.T., 1998,
MNRAS, 299, 83
\bibitem{cannizzo93}
Cannizzo, J.K., 1993, ApJ, 419, 318
%\bibitem{deMartino}
%de Martino, D., Buckley, D.A.H., Mouchet, M., Mukai, K., 1994, A\&A,
%284, 125
\bibitem{flebbeandall}
Flebbe, O., M\"unzel, S., Herold, H., Riffert, H., Ruder, H., 
1994, ApJ, 431, 754
\bibitem{} 
Godon, P. 1997, ApJ, 480, 329
\bibitem{GT}
Goldreich, P., Tremaine, S., 1979, ApJ, 233, 857
%\bibitem{HCM}
%Hellier, C., Cropper, M., Mason, K.O., 1991, MNRAS, 248, 233
\bibitem{EXHydrae}
Hellier, C., Mason, K.O., Rosen, S.R., Cordova, F.A., 1987, MNRAS,
228, 463
\bibitem{EXHYa2}
Hellier, C., Mason, K.O., Smale, A.P., Corbet, R.H.D., O'Donoghue, D.,
Barrett, P.E., Warner, B., 1989, MNRAS, 238, 1107
\bibitem{masahito}
Hirose, M., Osaki, Y., 1990, PASJ, 42, 135
\bibitem{notnowKato}
Kato, T., Nogami, D., 1997, PASJ, 49, 481
\bibitem{KL}
King, A.R., Lasota, J.-P., 1991, ApJ, 378, 674 
\bibitem{}
Kunze, S., Speith, R., Riffert, H., 1997, MNRAS, 289, 889
\bibitem{l89}
Lubow, S.H., 1989, ApJ, 340, 1064
\bibitem{l91}
Lubow, S.H., 1991, ApJ, 381, 259
\bibitem{Lubow_Ogilvie}
Lubow, S.H., Ogilvie, G., 1998, ApJ, 504, 983
\bibitem{marshhorne}
Marsh, T.R., Horne, K., 1990, ApJ, 349, 593
\bibitem{meg}
Meglicki, G., Wickramasinghe, D.T., Bicknell, G., 1993, MNRAS, 264, 691
%\bibitem{}
%Molteni, D. Belvedere, G., Lanzafame, G, 1991, MNRAS, 249, 748
\bibitem{}
Monaghan, J.J., 1992, ARAA, 30, 543
\bibitem{}
Murray, J.R., 1996a, MNRAS, 279, 402
\bibitem{}
Murray, J.R., 1996b, in ``Accretion phenomena and related outflows'',
eds  D.T. Wickramasinghe, G.V. Bicknell \& L. Ferrario, PASP, p 770
\bibitem{}
Murray, J.R., 1998, MNRAS, 297, 323
\bibitem{}
Murray, J.R., Armitage, P.J., 1998, MNRAS, in press
\bibitem{}
Norton, A.J., Hellier, C., Beardmore, A.P., Wheatley, P.J., Osborne,
J.P., Taylor, P., 1997, MNRAS, 289, 362
\bibitem{}
Papaloizou, J.C.B., Lin, D.N.C., 1995, ARAA, 33, 505
\bibitem{}
Patterson, J., 1994, PASP, 106, 209
%\bibitem{}
%Patterson, J., Kemp, J., Saad J., Skillman,  D.R.
%Harvey, D., Fried, R., Thorstensen, J.R., Ashley, R., 1997,
%PASP, 109, 468 
\bibitem{}
Riffert, H., Herold, H., Flebbe, O., Ruder, H., 1995,
Comp. Phys. Comm., 89, 1
\bibitem{} R\'o\.zyczka, M., Spruit, H.C., 1989, in ``Theory of Accretion
Disks'', eds F. Meyer, W.J. Duschl, J. Frank, E. Meyer-Hofmeister,
Kluwer, Dordrecht, p 341
\bibitem{}
Savonije, G.J., Papaloizou, J.C.B., Lin, D.N.C., 1994, MNRAS, 268, 13
\bibitem{}
Sawada, K., Matsuda, T., Hachisu, I., 1986, MNRAS, 221, 679
\bibitem{} 
Shakura, N.I., Sunyaev, R.A., 1973, A\&A, 24, 337
\bibitem{}
Spruit, H.C., 1987, A\&A, 184, 173
\bibitem{st}
Steeghs, D., Harlaftis, E.T., Horne, K., 1997, MNRAS, 290, L28
\bibitem{}
Tovmassian, G. et al., 1998, A\&A, 335, 227
%\bibitem{Sterken}
%Sterken, C., Vogt, N. Freeth, R., Kennedy, H.D., Marino, B.F., Page,
%A.A., Walker, W.S.G., 1983, A\&A, 118, 325
\bibitem{Warnerbook}
Warner, B., 1995, Cataclysmic Variable Stars. Cambridge University
Press, Cambridge
\bibitem{B&W}
Warner, B., Wickramasinghe, D.T., 1991, MNRAS, 248, 370
\bibitem{steveandall}
Watkins, S. J., Bhattal, A. S., 
Francis, N., Turner, J.A., Whitworth, A. P., 
1996, A\&AS, 119, 177
\bibitem{}
Whitehurst, R., 1988, MNRAS, 232, 35
\bibitem{graemy1}
Wynn, G., King, A.R., 1992, MNRAS, 255, 83
\bibitem{graemy}
Wynn, G., King, A.R., 1995, MNRAS, 275, 9
\bibitem{}
Yukawa, H., Boffin, H.M.J., Matsuda, T., 1997, MNRAS, 292, 321
\end{thebibliography}
\end{document}